\documentclass[aps, prb, amssymb, amsmath, superscriptaddress, twocolumn]{revtex4-1}

\usepackage{graphicx}
\usepackage{color}
\usepackage{amsmath}
\usepackage{amssymb}
\usepackage[colorlinks=true]{hyperref}

\begin{document}
	\title{Cyclotron reonance in a kagome spin liquid candidate material}
	
\author{Byungmin Kang}
\thanks{bkang119@mit.edu}
\affiliation{
Department of Physics, Massachusetts Institute of Technology, Cambridge, Massachusetts 02139, USA
}

\author{Patrick A. Lee}
\thanks{palee@mit.edu}
\affiliation{
Department of Physics, Massachusetts Institute of Technology, Cambridge, Massachusetts 02139, USA
}

\begin{abstract}
We propose cyclotron resonance as an optical probe for emergent fractionalized excitations in $\mathrm{U}(1)$ quantum spin liquids, focusing on kagome antiferromagnets. In contrast to conventional systems, where cyclotron resonance directly couples to charged carriers, spinons in spin liquids are charge-neutral and interact only through an emergent gauge field. We identify two key mechanisms by which an external physical electromagnetic field induces emergent electric and magnetic fields, enabling indirect coupling to spinons. Using these mechanisms, we compute the absorption rate of the cyclotron resonance response for Dirac spinons forming Landau levels. Our analysis shows that, although the absorption per layer is small, the absence of a skin-depth limitation in insulating spin liquids allows for cumulative absorption comparable to graphene in realistic sample sizes for the recently discovered spin-liquid candidate material YCu${}_3$(OH)${}_6$Br${}_2$[Br${}_{1-y}$(OH)${}_y$]. Our findings shows that cyclotron resonance is a viable experimental probe of spinon Landau quantization and emergent gauge fields, providing powerful positive experimental signatures of quantum spin liquids. 
\end{abstract} 

\maketitle

\section{Introduction}
A spin liquid is a quantum state of matter in which the bare electronic degrees of freedom exhibit fractionalization due to strong interactions~\cite{ANDERSON1973, Savary2016}. The usual mechanism behind this highly nontrivial state usually arises from magnetic frustration, which suppresses conventional ordering and instead gives rise to emergent, fractionalized excitations as the low-energy description. A strong candidate for realizing the spin liquid phase is the ground state of the antiferromagnetic kagome lattice~\cite{PhysRevB.45.12377, ran2007projected, PhysRevB.77.224413}. Due to geometric frustration, many theory studies~\cite{PhysRevB.45.12377, ran2007projected, PhysRevB.77.224413, RevModPhys.89.025003, yan2011spin, PhysRevLett.109.067201} suggest that kagome antiferromagnets can potentially realize either a $\mathbb{Z}_2$- or a $\mathrm{U}(1)$-spin liquid, where the latter exhibiting an emergent gapless gauge photon in addition to spinons. 

While the spin liquid is characterized by  the absence of conventional ordering, the search for the realization of a spin liquid phase in real materials has been hampered by the paucity of unambiguous \textit{positive} signatures. We are motivated by the recent discovery of the kagome spin-liquid candidate material YCu${}_3$(OH)${}_6$Br${}_2$[Br${}_{1-y}$(OH)${}_y$] (YCOB)~\cite{zheng2025unconventional, PhysRevX.15.021076, PhysRevB.109.L201104}, particularly the observation of magnetization plateau near 1/9 and oscillations in the vicinity of the plateau. These phenomena were interpreted as originating from  Dirac-like spinons which form Landau levels (LLs). Here, we briefly summarize the proposed picture. The low energy excitations in high magnetic field near the 1/9 magnetization is assumed to be described by fermionic spinon bands. The assumption of $2\pi/3$ flux per unit cell (since confirmed by projected variational Monte Carlo studies \cite{PhysRevLett.133.096501, kumar2025dirac}) gives rise to 9 bands. It is further assumed that bands 5 and 6 touch and produce Dirac nodes.\cite{kumar2025dirac} The Zeeman effect of a magnetic field $B_0$ places the chemical potential for the spin down spinon at the node and  and produces a V shaped 1/9 plateau, as seen in the experiment. If the magnetic field is further increased the Zeeman energy shifts the chemical potential away from the nodes and produces small Fermi pockets. At the same time, the $z$-component of the magnetic field gives rise to a gauge magnetic field $b$ which acts on the orbital motion of the spinons \cite{PhysRevB.109.L201104}, producing LL's. The Zeeman energy of the magnetic field $B$ moves the chemical potential across these LL's give rise to quantum oscillations as a function of $B$. In this paper, we aim to  obtain a direct probe of the LL structure formed by spinons using \textit{cyclotron resonance}~\cite{PhysRev.102.1030, dresselhaus2018solid}. Here we can apply a  magnetic field $B$ at or near $B_0$ that sets the 1/9 plateau. The orbital gauge magnetic field which is proportional to the $z$-component of $B$ will produce LL's that become resonant with an applied AC electromagnetic field. This resonance can be measured in the usual way by measuring the absorption of the AC field while scanning the magnitude or the orientation of $B$ or by changing the AC frequency. 

The cyclotron resonance, which corresponds to the resonant absorption of an electromagnetic field by charged particles subject to an external magnetic field, has been successfully used to reveal the LL structure of semiconductors, two-dimensional electron gas, and, notably, graphene~\cite{PhysRevLett.97.266405, PhysRevB.85.205407}. However, in our $\textrm{U}(1)$ spin liquid case, the LLs are formed by \textit{charge-neutral} spinons, which do not couple directly to the probing physical electromagnetic field, consistent with the fact that spin-liquid material is electrically insulating. Instead, spinons couple to an \textit{emergent} gauge field, and we focus on two mechanisms by which this emergent gauge field, particularly the emergent electric and magnetic field, can be induced by the physical electromagnetic field. Furthermore, compared with conventional cyclotron resonance, this approach offers a striking advantage: the skin-depth problem, namely, the exponential decay of electromagnetic field amplitude which limits the size of the signal in conventional metallic materials, is absent in the electrically insulating spin-liquid systems. Our analysis below indicates that the estimated absorption rate of cyclotron resonance per layer is small; however, this contribution can accumulate across layers, leading to realistic values for the recently discovered spin-liquid candidate material YCOB. We also note that, although our estimates are based on YCOB, our mechanism is broadly applicable, as it relies only on the presence of spinons coupled to an emergent $\textrm{U}(1)$ gauge field. We would like to point out the the idea of using cyclotron resonance as a probe for spin liquid phases was introduced earlier in   Ref.~\onlinecite{PhysRevB.100.155150}. In their work, the coupling arises from the emergent gauge magnetic response to the physical probing magnetic field via the Ioffe–Larkin mechanism~\cite{PhysRevB.39.8988}, which is applicable primarily to small band-gap insulators---wheras our mechanisms are applicable for a large bandgap insulator. Furthermore, their analysis focuses on spin-liquid systems with a spinon Fermi surface, whereas our work targets $\textrm{U}(1)$ spin liquids, where the spinons exhibit Dirac dispersion and form Landau levels under an emergent gauge magnetic field. Therefore, the two works address complementary regimes of applicability in probing quantum spin liquids via cyclotron resonance.

\section{Kagome Spin Liquid and Emergent Gauge Field}
The kagome antiferromagnet is described by the following microscopic Hamiltonian: 
\begin{equation}
\label{eq:kagome-ham}
H = \sum_{\langle i, j \rangle} \big( J \boldsymbol{S}_i \cdot \boldsymbol{S}_j + \boldsymbol{D}_{i, j}\cdot \boldsymbol{S}_i \times \boldsymbol{S}_j \big) - h \sum_i S_i^z, 
\end{equation}
where $J$ is the antiferromagnetic coupling and $\boldsymbol{D}_{i,j}$ is the Dzyaloshinskii-Moriya (DM) vector (for which we focus only on the $z$-component and  denote its amplitude by $D$ in this paper), and $h$ is the external magnetic field along the $z$-direction. Following recent experiment~\cite{zheng2025unconventional, PhysRevX.15.021076}, the system at $1/9$ filling exhibits $\textrm{U}(1)$ spin-liquid behavior, suggesting an emergent spinons with Dirac-like dispersion. Being charge-neutral particles, spinons do not couple to the physical electromagnetic field; instead, they couples to an emergent gauge field which supports gapless excitations. Below, we outline key features of the $\textrm{U}(1)$ spin liquid that are relevant to our cyclotron resonance framework. 

The low-energy theory of spinons is captured by $N_f$ copies of two-component Dirac particles: 
\begin{equation}
\label{eq:dirac-spinon-ham}
\hat{H}_0 = \sum_{\alpha = 1}^{N_f} \sum_{\boldsymbol{k}} \hbar v_D \boldsymbol{f}_{\boldsymbol{k}; \alpha}^\dagger [\boldsymbol{k} \cdot \boldsymbol{\tau}_\alpha] \boldsymbol{f}_{\boldsymbol{k}; \alpha} ,  
\end{equation}
where $v_D$ is the Dirac velocity, assumed to be identical to all nodes, $\alpha$ labels each Dirac node, the momentum $\boldsymbol{k} = (k_x, k_y)$ is measured relative to the corresponding Dirac node, $\boldsymbol{f}^\dagger_{k; \alpha} = (f^\dagger_{k, \uparrow; \alpha}, f^\dagger_{k, \downarrow; \alpha})$ with $\uparrow$/$\downarrow$ denoting two components of each Dirac spinon, and $\boldsymbol{\tau} = (\sigma_x, \sigma_y)$ is a collection of two Pauli matrices acting on the two-component structure of each Dirac node. In addition to the spinons, the $\textrm{U}(1)$ spin liquid has a gapless emergent gauge field $\bold{a}$, which arises from the phase factor of the hopping expectation value between neighboring sites $j$ and $k$: $\langle \boldsymbol{f}_j^\dagger \boldsymbol{f}_k \rangle \approx \chi e^{\frac{i}{\hbar} \textrm{a}_{j k}^{(\textrm{d})}}$, where $\chi \sim 0.1$ is the modulus and $\textrm{a}_{j k}^{(\textrm{d})} = \int_{\boldsymbol{r}_j}^{\boldsymbol{r}_k} d\boldsymbol{l} \cdot \bold{a}(r)$ is the discrete emergent gauge field. By setting the emergent gauge scalar potential to $0$, the emergent electric and magnetic field are expressed as: 
 \begin{equation}
\begin{cases}
\bold{e} = - \dot{\bold{a}} \\ 
b = \nabla \wedge \bold{a} = \partial_x \bold{a}_y - \partial_y \bold{a}_x 
\end{cases} , 
\end{equation}
where the emergent electric field $\bold{e} = (\textrm{e}_x, \textrm{e}_y)$ consists of two in-plane components, and the emergent magnetic field $b$ is a single out-of-plane component oriented along the $z$-direction. Similar to photon, the emergent gauge field supports a propagating gapless gauge boson with velocity $v_g$, which is proportional to $J a/\hbar$, where $J$ denotes the Heisenberg coupling strength and $a$ is the lattice constant. As an oscillating gauge field $\bold{a} = \bold{a}_0 e^{i (\boldsymbol{q} \cdot \boldsymbol{r} - \omega t)}$, the emergent magnetic field is given by 
\begin{equation}
b = i (\boldsymbol{q} \wedge \bold{a}_0) e^{i (\boldsymbol{q} \cdot \boldsymbol{r} - \omega t)}  =  i \omega v_g (\hat{\boldsymbol{q}} \wedge \bold{a}_0) e^{i (\boldsymbol{q} \cdot \boldsymbol{r} - \omega t)}
\end{equation}
and the emergent electric field is given by 
\begin{equation}
\bold{e} = - \dot{\bold{a}} = i \omega \bold{a}_0 e^{i (\boldsymbol{q} \cdot \boldsymbol{r} - \omega t)} . 
\end{equation} 
Finally, the spinons couple to the emergent gauge field through minimal coupling: 
\begin{equation}
\label{eq:dirac-spinon-a-ham}
\hat{H}[\bold{a}] = \sum_{\alpha = 1}^{N_f} \sum_{\boldsymbol{k}} \hbar v_D \boldsymbol{f}_{\boldsymbol{k}; \alpha}^\dagger \Big[ \Big(\boldsymbol{k} + \frac{1}{\hbar} \bold{a} \Big) \cdot \boldsymbol{\tau}_\alpha \Big] \boldsymbol{f}_{\boldsymbol{k}; \alpha} . 
\end{equation}

\section{Perturbing Spin Liquid via external physical gauge field}
\label{sec:mechanisms}
In the following, we review two mechanisms by which external electromagnetic field generates an emergent gauge field. Specifically, we explain how an external electric (magnetic) field induces an gauge electric (magnetic) field through two distinct mechanisms. The first mechanism, which generates an emergent electric field from an external physical electric field, arises from magnetoelastic coupling~\cite{PhysRevB.87.245106}. The second mechanism, responsible for producing an emergent magnetic field in response to an external magnetic field, originates from the accumulated Berry phase~\cite{wen1989chiral, PhysRevB.46.5621, PhysRevB.109.L201104}. We note that the magnetoelastic coupling also introduces a coupling between the external electric field and the spin-current; however, this contribution is considerably weaker than the induced electric field. 

\subsection{Emergent Electric Field}
The first mechanism involves the generation of an emergent electric field by an external physical electric field. While details can be found in Ref.~\onlinecite{PhysRevB.87.245106}, here we summarize and highlight key aspects important for our application. This process is enabled by the magnetoelastic response to the external electric field, where the charged lattice atoms undergo deformation in response to an in-plane external electric field. In the kagome spin liquid, the deformation of atoms introduces the modulation of Heisenberg and DM interaction, where the former generates the direct coupling between the emergent electric field and the external electric field and latter generates the direct coupling between the spin-current of spinon and the external electric field. It turns out that the effects from the spin-current is much weaker than that from the emergent electric field~\cite{PhysRevB.87.245106}, therefore, we mostly focus on the effects from the emergent electric field. 

The modulation in the Heisenberg term by the in-plane external electric field $\boldsymbol{E}_\parallel$ is given by 
\begin{align}
\label{eq:e-E-coupling-Ham}
\delta H &= \sum_{\langle i,j \rangle} \delta J_{i,j} \boldsymbol{S}_i \cdot \boldsymbol{S}_j = \sum_{\langle i,j \rangle} \frac{\delta J_{i,j}}{\delta u} \frac{\delta u}{\delta E} E \, \boldsymbol{S}_i \cdot \boldsymbol{S}_j \nonumber \\
&= \frac{e}{k_{\textrm{Cu}}} n_\triangle \int d^2 r \boldsymbol{E}_\parallel \cdot \bold{e} , 
\end{align}
where $a$ is the lattice constant of the kagome lattice, $n_\triangle = \frac{1}{\sqrt{3} a^2}$ is the density of triangles, $k_{\textrm{Cu}} a^2 \sim 1 \textrm{eV}$ is the spring constant of Cu atom,  $\delta J_{i,j}$ is the modulated Heisenberg coupling via the magnetoelastic mechanism, $e$ is the electric charge, and $\bold{e}$ is the emergent electric field. From the Dirac-like spinons coupling to an emergent gauge field Eq.~\eqref{eq:dirac-spinon-a-ham}, one can employ the random phase approximation (RPA) to calculate the linear response of the emergent gauge electric field $\bold{e}$ to an external in-plane electric field $\boldsymbol{E}_\parallel$~\cite{PhysRevB.87.245106}: 
\begin{equation}
\label{eq:e-from-E}
\bold{e} = \frac{\omega}{\sigma_D} \frac{e}{k_{\textrm{Cu}}} n_\triangle \boldsymbol{E}_\parallel := \chi_{\textrm{e}} e \boldsymbol{E}_\parallel
\end{equation}
where $\sigma_D = N_f/(16 \hbar)$. Note that $\chi_{\textrm{e}}$ has the frequency ($\omega$) dependence. 

Similarly, the external in-plane electric field also modulates the DM interaction, thereby inducing a coupling between the external electric field and the spinon spin current~\cite{PhysRevB.87.245106}: 
\begin{equation}
\delta H_{DM} = \frac{2 e}{K_\textrm{eff} a} \frac{D \chi^f}{J} \int d^2 \boldsymbol{r} \, \boldsymbol{E}_\parallel \cdot (\hat{z} \times \boldsymbol{j}_{S^z} ) , 
\end{equation}
where $K_{\textrm{eff}} = \big( \frac{1}{K_{\textrm{O}} \cos \alpha} - \frac{1}{K_{\textrm{Cu}}} \big)^{-1}$ is the effective spring constant from $\textrm{Cu}^{2+}$ and $\textrm{O}^{2-}$ configuration with $\alpha$ being the Cu-O bond angle and $\boldsymbol{j}_{S^z}$ is the $z$-component spin current. However, due to the presence of the $D/J$ factor in the coupling, typically $\sim 0.1$ in most spin-liquid materials, the conductivity contribution arising from DM modulation carries an overall $(D/J)^2$ factor compared to that from $J$ modulation, resulting in an effect of only about $1$ \%, as noted in Ref.~\onlinecite{PhysRevB.87.245106}. Therefore, we omit the contribution from DM modulation in the following discussion. 

\subsection{Emergent Magnetic Field}
The second mechanism involves the generation of an emergent magnetic field by an external physical magnetic field. It is well known that the gauge magnetic field, equivalently, the magnetic flux, is determined by the Berry phase associated with fermion hopping around a plaquette~\cite{wen1989chiral, PhysRevB.46.5621, PhysRevB.109.L201104}. According to the recent estimates~\cite{PhysRevB.109.L201104} for the emergent gauge magnetic field in the kagome spin liquid system [Eq.~\eqref{eq:kagome-ham}] can be induced by the external magnetic field via 
\begin{equation}
\label{eq:b-from-B}
b = \chi_b \frac{e}{c} B_\perp , 
\end{equation}
where $\chi_b \sim -10$ in the case of a static $B_\perp$~\cite{PhysRevB.109.L201104} around the $1/9$ plateau. In cyclotron resonance, however, the probe is an oscillating AC field giving oscillatory perturbation $\delta B_\perp$ and thus corresponds to a Floquet drive rather than a static perturbation. In this case, the induced response $\delta b$ also oscillates at the same frequency, though the effective $\chi_b$ may differ from its static value.

\section{Cyclotron Resonance Absorption Rate}
In this section, we investigate how cyclotron resonance can be employed to optically probe the spinon spectrum, specifically focusing on the Landau levels (LLs) formed by Dirac spinons, via the two mechanisms discussed in the previous section. As explained earlier, these mechanisms describe how an emergent electric (magnetic) field is induced by an external physical electric (magnetic) field. Therefore, the coupling between the external probing field and spinons is indirect, as spinons couple directly only to the emergent gauge field. 

In the following, we restrict our attention to a single valley degrees of freedom, corresponding to a single $\alpha$ in Eq.~\eqref{eq:dirac-spinon-ham}; later, we include the appropriate degeneracy factor to recover the full expression. Moreover, we assume that the spin-liquid material consists of a single two-dimensional layer. As in the conventional cyclotron resonance setting, we assume the presence of a background magnetic field, responsible for the formation of the LLs, along with an additional probe electromagnetic field that perturbs the system. In our kagome spin liquid setting, this corresponds to working at $1/9$-filling, where spinon LLs are formed. Therefore, we decompose the emergent gauge field $\bold{a}$ as $\bold{a}_\textrm{bkg} + \bold{a}_\textrm{opt}$, where $\bold{a}_\textrm{bkg}$ denotes the the background emergent gauge field responsible for the LL formation, and $\bold{a}_\textrm{opt}$ represents the perturbing emergent gauge field induced by the optical probing field. Below, we compare the cyclotron resonance in kagome spin-liquid systems with that in graphene, closely following the analysis for graphene, e.g., in Refs.~\onlinecite{PhysRevLett.97.266405, PhysRevB.85.205407}, but adapting it to the spin-liquid context. 

In the presence of the background emergent gauge magnetic field, the spinons form the Landau levels (LLs) with eigenenergies: 
\begin{equation}
E_{n, \pm} = \pm \hbar \omega_\textrm{LL} \sqrt{n} ,
\end{equation}
where $n = 0, 1, 2, \ldots$ labels the Landau level, $\omega_\textrm{LL} = \sqrt{2} v_D / l_b$ denotes the characteristic cyclotron frequency with $l_b = \sqrt{\hbar/b}$ being the magnetic length, and the corresponding eigenvectors are: 
\begin{equation}
\psi_{n, k, \pm} = d_n \left( \begin{array}{c}
(1 - \delta_{n, 0}) \phi_{n-1, k} \\
\pm i \phi_{n, k}
\end{array} \right) , 
\end{equation}
where $d_n = (1 - \frac{1}{\sqrt{2}}) \delta_{n, 0} + \frac{1}{\sqrt{2}}$. By choosing the Landau gauge $\bold{a}_\textrm{bkg} = (0, b x, 0)$, 
\begin{equation}
\phi_{n, k} (x, y) = \frac{1}{\sqrt{2^n n! \sqrt{\pi}}} \frac{1}{\sqrt{L_x L_y}} e^{i k y} e^{-\frac{(x-kl_b^2)^2}{2 l_b^2}} H_n \Big[ \frac{x-k l_b^2}{l_b} \Big] 
\end{equation}
where $k$ is the momentum along the $y$-direction and $H_n$ is the $n$th Hermite polynomial. 

\begin{figure}
\includegraphics[width=0.45\textwidth]{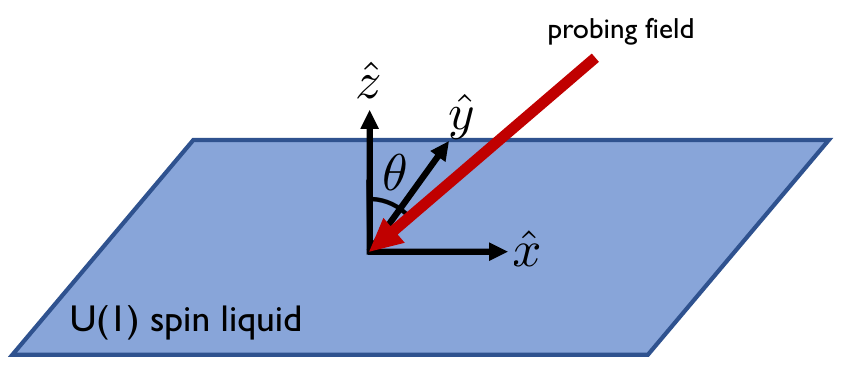}
\caption{Probing field enters spin liquid material with an angle $\theta$, where spin liquid material lies on the $xy$-plane and the probing field lies on the $xz$-plane.}
\label{fig:light-spin-liquid}
\end{figure}

Under the perturbing gauge field $\bold{a}_\textrm{opt}$, which is induced from the optical probing AC field, the corresponding term in the Hamiltonian becomes 
\begin{equation}
\delta H = \hbar v_D \boldsymbol{\tau} \cdot \frac{\bold{a}_\textrm{opt}}{\hbar} = v_D e^{i \boldsymbol{q} \cdot \boldsymbol{r}} \left( \begin{array}{cc}
0 & \textrm{a}_x - i \textrm{a}_y \\ 
\textrm{a}_x + i \textrm{a}_y & 0 
\end{array}
\right) , 
\end{equation}
where we separate the spatial dependence into $e^{i \boldsymbol{q} \cdot {r}}$ and the time dependence into $(\textrm{a}_x, \textrm{a}_y) = (\textrm{a}_x (t), \textrm{a}_y (t))$. Since we are mainly interested in the low-energy regime, we set $\boldsymbol{q} = 0$ from this point onward. Following the discussion in Sec.~\ref{sec:mechanisms}, we consider $\bold{a}_{\textrm{opt}}$ corresponding to induced gauge electric and magnetic field. Care must be taken because the emergent electric field is induced by an \textit{in-plane} external electric field, and the emergent magnetic field is induced by an \textit{out-of-plane} external magnetic field. Therefore, we specifically focus on the in-plane electric field component and the out-of-plane magnetic field component of the probing electromagnetic field. To this end, consider a circular polarized light incident on the sample at an angle $\theta$, as illustrated in FIG.~\ref{fig:light-spin-liquid}. The in-plane physical electric field is then given by $\boldsymbol{E}_{\parallel} = \textrm{Re} \big[ E_0 (\cos \theta \hat{x} \pm i \hat{y} ) e^{i(\boldsymbol{q} \cdot \boldsymbol{r} - \omega t)} \big]$ and the out-of-plane component physical magnetic field is given by $\boldsymbol{B}_{\perp} = \textrm{Re} \big[ \mp i E_0 \sin \theta \hat{z} e^{i(\boldsymbol{q} \cdot \boldsymbol{r} - \omega t)} \big]$, where $\pm$ denotes the right-/left-handed polarization. These fields induce the emergent electric field according to Eq.~\eqref{eq:e-from-E} and emergent magnetic field according to Eq.~\eqref{eq:b-from-B}. 

Below, we evaluate the cyclotron resonance contributions separately from the induced emergent electric field and the induced magnetic field contribution. For a positively circularly polarized light, the corresponding matrix element from the emergent AC electric field is given by 
\begin{align}
&|\langle \psi_{m, k', \alpha} \vert \delta H \vert \psi_{n, k, \alpha} \rangle | \nonumber \\
&= \frac{v_D \textrm{e}_0}{\omega} d_n d_m \Big( \frac{1 + \cos \theta}{\sqrt{2}} \delta_{m, n-1} + \frac{1 - \cos \theta}{\sqrt{2}} \delta_{m-1, n} \Big) , 
\end{align}
where $\textrm{e}_0 (=\chi_e e E_0)$ is the amplitude of the emergent electric field.  If we focus on $n \to n+1$ transition and assume that the $n$th LL is fully occupied and $(n+1)$th LL is completely empty, the Fermi's golden rule gives the following transition rate per unit area: 
\begin{align}
&R_{n \to n+1}^{\textrm{e}} = \frac{1}{L_x L_y} \sum_k \frac{2\pi}{\hbar} |\langle n+1, \alpha, k' \vert \delta H \vert n, \alpha, k \rangle |^2 \nonumber \\ 
&\qquad \qquad \qquad \times \delta \big(\hbar \omega - (E_{n+1} - E_n ) \big) \nonumber \\ 
&\mapsto \frac{v_D^2 e^2 \chi_e^2 E_0^2 (1 + \cos \theta)^2}{2 (l_b)^2 \hbar \omega^2} d_n^2 d_{n+1}^2 \frac{1}{\pi} \frac{\hbar \tau^{-1}}{(\hbar \omega - \Delta E)^2 + (\hbar \tau^{-1})^2} , 
\end{align}
where we introduced characteristic time $\tau$ to include the peak broadening due to disorder. We note that if linearly polarized light is used, with the probing AC electric field oriented parallel to the $y$-direction in Fig.~\ref{fig:light-spin-liquid}, the resulting transition rate is similar in magnitude but becomes independent of the incident angle. Given the transition rate $R$, the absorption coefficient is given by~\cite{chuang1996physics, PhysRevB.85.205407} 
\begin{equation}
\alpha_{\textrm{abs}} = \frac{R}{S/ \hbar \omega} , 
\end{equation}
where $S = \frac{n_r c}{8\pi} E_0^2$ is the Poynting vector of the probing electromagnetic field. Due to disorder-induced broadening, the absorption rate exhibits peaks near the resonant transitions. The integrated absorption rate around a peak, which carries a dimension of energy, can be extracted from experimental data. To obtain a realistic estimate of the integrated absorption rate for our spin liquid system, we compare it to the case of graphene. For graphene, the integrated absorption rate has the energy scale $I_\textrm{graphene} = \frac{4\pi}{n_r c} \frac{e^2}{\hbar} \hbar \omega_{\textrm{graphene}}$ in cgs units (or equivalently $\frac{1}{n_r \epsilon_0 c} \frac{e^2}{\hbar} \omega_{\textrm{graphene}}$ in SI units), where $\omega_{\textrm{graphene}}$ denotes the energy scale of the Landau level gap in graphene. In our case, the integrated absorption rate is given by: 
\begin{equation}
I_\textrm{spinon}^{\textrm{e}} = \frac{4\pi}{n_r c} \frac{N_f}{4} \frac{e^2}{\hbar} (\chi_e)^2 \frac{(1 + \cos \theta)^2}{2} \hbar \omega_{\textrm{LL}} , 
\end{equation}
where we now recover the number of Dirac nodes $N_f$ in the expression. Therefore, we compute the following ratio: 
\begin{align}
\label{eq:I-ratio-e}
&I_\textrm{spinon}^{\textrm{e}}/I_\textrm{graphene} = \frac{N_f}{4} (\chi_e)^2 \frac{(1 + \cos \theta)^2}{2} \frac{\hbar \omega_\textrm{LL}}{\hbar \omega_{\textrm{graphene}}} \nonumber \\ 
&= \frac{N_f}{4} \Big(\frac{16 \hbar \omega_\textrm{LL}}{N_f \sqrt{3} k_{\textrm{Cu}} a^2}\Big)^2 \frac{(1 + \cos \theta)^2}{2} \frac{\omega_\textrm{LL}}{\omega_{\textrm{graphene}}} . 
\end{align}
Based on the recent experiments on YCOB~\cite{zheng2025unconventional}, we can estimate value of the ratio. Setting $N_f = 3$, $k_{\textrm{Cu}} a^2 = 1$ eV, and the LL gap $\hbar \omega_{\textrm{LL}} \sim 1$ meV (which is $1/10$ of the typical $\hbar \omega_{\textrm{graphene}} \sim 10$ meV), the ratio in Eq.~\eqref{eq:I-ratio-e}, aside from the angular dependence, gives $\sim 
7 \times 10^{-7}$. In the conventional materials probed by cyclotron resonance, the number of layers accessible to the probe is limited by the penetration depth, a constraint known as the \textit{skin-depth phenomenon} in metals. In contrast, our spin liquid material is electrically insulating, eliminating the skin-depth limitations. Consequently, for $\sim 10^6$ layers along the $c$-direction, corresponding to a sample thickness of roughly 1 mm, the absorption intensity reaches that of graphene, which is $\sim 1 \%$, a value that has been experimentally observed~\cite{PhysRevLett.97.266405, PhysRevB.85.205407}. 

Next we perform the same calculation for the contribution from the emergent magnetic field. Note that this involves a magnetic dipole transition which is much weaker than the electric dipole transition. On the other hand, the gauge magnetic field is comparable in amplitude to the physical magnetic field~\cite{PhysRevB.109.L201104}, in contrast to the small value of the gauge electric field. Therefore these two mechanisms need to be compared.  We obtain the transition rate due to the AC magnetic field as:
\begin{align}
&R_{n \to n+1}^b = \frac{v_D^2 v_g^2 b^2}{l_b^2 \hbar \omega^2} d_n^2 d_{n+1}^2 \frac{1}{\pi}\frac{\hbar \tau^{-1}}{(\hbar \omega - \Delta E)^2 + (\hbar \tau^{-1})^2} \nonumber \\
&= \frac{v_D^2 v_g^2 \chi_b^2 \big( \frac{e}{c} \big)^2 (E_0 \sin \theta)^2}{l_b^2 \hbar \omega^2} d_n^2 d_{n+1}^2 \frac{1}{\pi}\frac{\hbar \tau^{-1}}{(\hbar \omega - \Delta E)^2 + (\hbar \tau^{-1})^2} 
\end{align}
and the integrated absorption rate
\begin{equation}
I_\textrm{spinon}^b = \frac{4\pi}{n_r c} \frac{N_f}{4} \frac{e^2}{\hbar} \big( \frac{v_g}{c} \big)^2 \chi_b^2 \sin^2 \theta \hbar \omega_{\textrm{LL}} . 
\end{equation}
Note that the $\sin^2 \theta$ factor would still survive if linearly polarized light is used, with the probing AC electric field oriented parallel to the $y$-direction in Fig.~\ref{fig:light-spin-liquid}. Compared with the integrated absorption rate of graphene, we get the following ratio: 
\begin{equation}
\label{eq:I-ratio-b}
I_\textrm{spinon}^{b}/I_\textrm{graphene} = \frac{N_f}{4} \big( \frac{v_g}{c} \big)^2 (\chi_b)^2 \sin^2 \theta \frac{\omega_\textrm{LL}}{\omega_{\textrm{graphene}}} . 
\end{equation}
Using $J \sim 4$ meV in YCOB~\cite{zheng2025unconventional}, which yields $v_g = Ja/\hbar \sim 10^4$ m/s, and $|\chi_b| \sim 10$, the estimated value of the ratio, aside from the angular dependence, is approximately $10^{-8}$, which is smaller but comparable to $I_\textrm{spinon}^{\textrm{e}}/I_\textrm{graphene}$ estimates. Note that one can disentangle the contributions of the two mechanisms from the absorption ratio data using the angular dependence given by Eqs.~\eqref{eq:I-ratio-e} and~\eqref{eq:I-ratio-b}.

\section{conclusion}
In this work, we have proposed cyclotron resonance as an optical probe for the spinon spectrum in $\textrm{U}(1)$ spin-liquid systems, focusing on the Landau levels (LLs) formed by Dirac spinons. Unlike conventional materials, where charged carriers couple directly to the physical electromagnetic field, spinons are charge-neutral and couple instead to an emergent gauge field. We have discussed two mechanisms by which the emergent gauge electric and magnetic fields can be induced by an external probing physical electromagnetic field: (i) via an in-plane electric field and (ii) via an out-of-plane magnetic field.

Building on these mechanisms, we have derived the cyclotron resonance response of spinons, highlighting its analogy to graphene while emphasizing key distinctions arising from the emergent gauge structure. Our analysis shows that, although the absorption rate per layer is small, the absence of the skin-depth limitation in electrically insulating spin-liquid materials allows for substantial accumulation over multiple layers. For realistic sample thicknesses, the integrated absorption can reach experimentally accessible levels.

Needless to say, the observation of cyclotron resonance-like peaks in an insulator like YCOB will be a direct confirmation for the interpretation of the magnetization structures as due to quantum oscillations~\cite{zheng2025unconventional}, and provide direct evidence for emergent fermions as well as gauge electric and magnetic fields. As our analysis applies broadly to $\textrm{U}(1)$ spin liquid materials, these results demonstrate that cyclotron resonance offers a viable route for probing spinon Landau quantization in conjunction with emergent gauge dynamics, providing a powerful experimental signature of spin liquids.

\section*{Acknowledgements} 

PL acknowledges support by DOE (USA) office of Basic Sciences Grant No. DE-FG02-03ER46076.

\bibliography{ref}

\end{document}